\documentclass[aps,prb,showpacs,twocolumn,superscriptaddress,letterpaper]{revtex4}
\usepackage{amsmath}
\usepackage{amssymb}
\usepackage{amsfonts}
\usepackage{graphicx}% Include figure files
\usepackage{array}
\newcommand{\PreserveBackslash}[1]{\let\temp=\\#1\let\\=\temp}
\newcolumntype{C}[1]{>{\PreserveBackslash\centering}p{#1}}
\newcolumntype{R}[1]{>{\PreserveBackslash\raggedleft}p{#1}}
\newcolumntype{L}[1]{>{\PreserveBackslash\raggedright}p{#1}}

\def\be{\begin{equation}}       \def\ee{\end{equation}}
\def\bea{\begin{eqnarray}}      \def\eea{\end{eqnarray}}

\begin{document}
%\title{The influence of interaction parameters on the Pairing Symmetry in Iron-based Superconductors}
\title{ Relationship between Pairing Symmetries and Interaction Parameters in Iron-Based Superconductors from Functional Renormalization Group Calculations}
\author{Jing Yuan}
\affiliation{Institute of Physics, Chinese Academy of Sciences, Beijing 100190, China}
%\author{Yupeng Wang}
%\affiliation{Institute of Physics, Chinese Academy of Sciences, Beijing 100190, China}
%\affiliation{Collaborative Innovation Center of Quantum Matter, Beijing, China}
\author{Jiangping Hu}
\affiliation{Institute of Physics, Chinese Academy of Sciences, Beijing 100190,
China}
\affiliation{Department of Physics, Purdue University, West Lafayette, Indiana 47907, USA}
\affiliation{Collaborative Innovation Center of Quantum Matter, Beijing, China}

\date{\today}

\begin{abstract}
Pairing symmetries of iron-based superconductors  are investigated systematically in a five-orbital model within the different regions of interaction parameters by functional renormalization group(FRG). Even for a fixed Fermi surface with both hole and electron pockets,  it is found that depending on interaction parameters,  a variety of pairing symmetries, including two types of $d$-wave  and two types of $s$-wave pairing symmetries,  can emerge.  Only the $d_{x^2-y^2}$ and the $s\pm$ waves are robustly supported by  the nearest-neighbor(NN)  intra-orbital   $J_1$ and the next-nearest-neighbor(NNN) intra-orbital    $J_2$  antiferromagnetic(AFM)  exchange couplings respectively. This study suggests that  the accurate initial input of interaction parameters are essential to make  FRG  an useful method to determine the leading channel of superconducting instability.\end{abstract}


\maketitle

\section{Introduction}

The discovery of high temperature Fe-based superconductors(FeSCs) is a major breakthrough in condensed matter physics field\cite{Hosono2008}. Since their great application prospect, lots of theoretical and experimental researches have been devoted to study the FeSCs \cite{Hirschfeld2011,Chubukov-2012,Scalapino-2012,Aswathy-2010,Stewart-2011,Johnston-2010,Richard-2015}. There is an increasing diversity of superconducting materials and more complicated characteristics. Nevertheless, the mechanism of FeSCs has not been confirmed thoroughly.

Iron-pnicitides usually contain both hole and electron pockets which are separated with momentum $(\pi,\pi)$ in 2-Fe Brillouin zone(BZ). Theoretically,   $s\pm$-wave\cite{Mazin2008,Bernevig2008,Kuroki2008,Hirschfeld2011}, which  displays a sign change between the superconducting orders on hole pockets and  electron pockets, is the most promising candidate.  However, once the Fermi surfaces change, $s\pm$ is not the leading instability in many theoretical studies.  For example, in the FeSCs with only electron pockets, such as  most iron-chalcogenides\cite{Chenxiaolong2010,FengKFeSe2011,ZhouFeSe2012}, or only hole pockets \cite{Sato2009,Yoshida2011,Reid2012}, the $d$-wave pairing channel is favored in most theoretical studies based on spin fluctuation mechanism and standard weak coupling treatments\cite{Thomale2011,Maiti2011,Maiti2012,Reid2012}. In addition,  orbital dependent sign change $s$-wave\cite{Hu-S4,cotliar},  $s^h_{\pm}$\cite{Tafti2013,Watanabe2014,Maiti2012} with sign-reversal between hole pockets,  $\eta$-pairing $s$-wave\cite{Hu-oddpairty} and   time-reversal symmetry breaking  states including $s+id$ state\cite{Khodas2012,Platt2012} and $s+is$ state\cite{Maiti2013,Marciani2013}  have been proposed theoretically.

While most weak coupling approaches have targeted  on the  change of pairing symmetries depending on the variations of Fermi surfaces,    the relationship between interaction parameters and pairing symmetries has not been well explored. In our paper based on functional renormalization group(FRG) \cite{Shankar1994,Thomale2013,Metzner2012,Honerkamp2010,Wang2009-1}, an unbiased weak coupling approach, we  systematically investigate the effect of interaction  parameters  on pairing symmetries  in a fixed Fermi  surface topology that exhibits both hole and electron pockets \cite{Graser2010}.  We consider interactions including intra-orbital Coulomb coupling $U$, inter-orbital Coulomb coupling $U'$, Hund's coupling $J_H$, pairing hopping term $J_{pair}$, nearest-neighbour(NN)  antiferromagnetic(AFM)  exchange coupling $J_1$, and next-nearest-neighbour(NNN) AFM exchange coupling $J_2$.  We calculate the phase diagrams of the leading superconducting instability. It is found that depending on interaction parameters, all $s\pm$, $d_{x^2-y^2}$, $d_{xy}$, and $s^h_\pm$ phases  can be emerged within the reasonable interaction regions. This result suggests that the theoretical predicting power from this current theoretical method is rather limited as it is difficult to extract  accurate effective microscopic interaction parameters for   complex systems such as iron-based superconductors.

\section{Method and Model}
We use the FRG method to analyze the pairing symmetry with different types of  interactions. FRG is a numerical calculation method for weak to moderate electron correlation systems. As it takes into account all virtual two-electron scattering processes and  calculates all the electronic instabilities without bias, FRG is considered to be a resultful method in calculating the instabilities and pairing symmetries of materials.

The results of FRG  are known to be sensitive to  Fermi surface topology and the details of Fermi surfaces properties.  Moreover, these results also closely rely on the initial type and value of interactions, which is the central focus of this paper.   For the sake of convenience of numerical calculation, we discretize the momenta by  dividing  the BZ into $N$ patches\cite{Zanchi2000,Thomale2009}, here $N=80$, and each patch contains one Fermi surface segment. So in the numerical process, we treat the particle momentum with the patch index. Approximately, we regard the coupling function as a constant in each patch and  ignore the frequency dependence of the vertex function and the self-energy\cite{Honerkamp2001,Wang2009-2}. More specifically, FRG mainly outputs the effective interaction as the momentum continually tends to the Fermi surface. The form of four-particle effective interaction is
\begin{eqnarray}
V_\Lambda(k_1,k_2,k_3,k_4)c_{k_1b_1\alpha}^{\dagger}c_{k_2b_2\beta}^{\dagger}c_{k_4b_4\beta}c_{k_3b_3\alpha},
\end{eqnarray}
where $V(k_1,k_2,k_3,k_4)$ exhibits a four-point interaction vertex with the incoming and outgoing  momenta $k_1,k_2,k_3,k_4$, here we label these particles with the discrete Fermi surface patches; $b_1,b_2,b_3,b_4$ denote the corresponding  band indexes of the four particles and $\alpha,\beta$ are spin indexes. The  energy cutoff $\Lambda$ is the FRG flow parameter which makes the energy finally approach around the Fermi surface. Apparently, for the superconducting channel $\mathbf{k_1}=-\mathbf{k_2}=\mathbf{k}$, $\mathbf{k_3}=-\mathbf{k_4}=\mathbf{p}$ and the  four-point function becomes  $V_{SC}(\mathbf{k},-\mathbf{k},\mathbf{p},-\mathbf{p})$. We rewrite the function into eigen-mode,
\begin{equation}
V_{\Lambda,SC}(\mathbf{k},-\mathbf{k},\mathbf{p},-\mathbf{p})=\sum_iw_i(\Lambda)f^*_i(\mathbf{k})f_i(\mathbf{p}),
\label{V}\end{equation}
where $i$ is the decomposition index. The leading instability of the superconducting channel corresponds to the eigenvalue $w_1(\Lambda)$ which is the most diverging one as the decreasing of $\Lambda$, and the homologous form factor $f_1(k)$ tells us the detailed information of superconducting order parameter, the pairing symmetry and the gap structure.

\begin{figure}
  \centering
  \includegraphics[width=8.8cm]{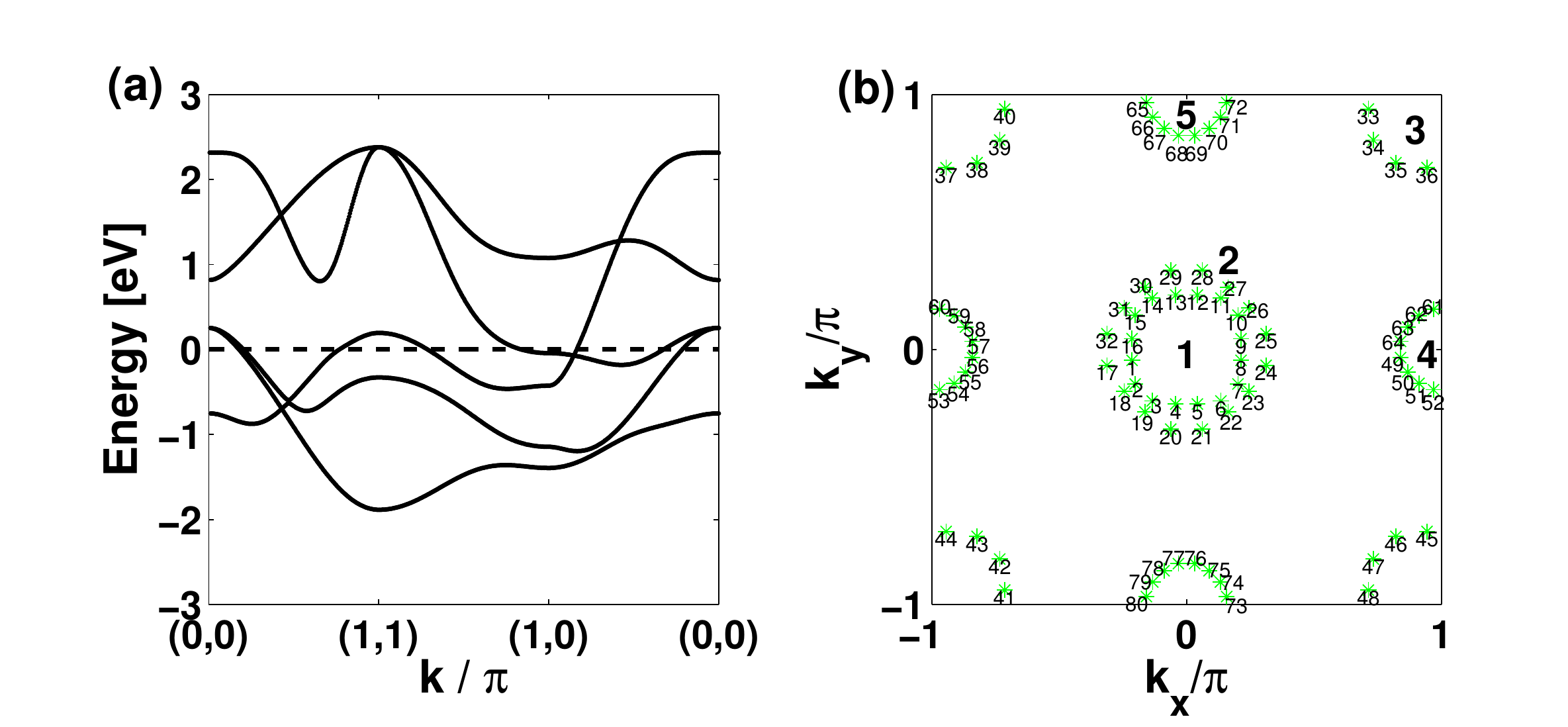}\\
  \caption{(color online)(a) The band structure of the five-orbital tight-binding model with chemical potential $\mu=-0.20$ with  0.317 hole doping per Fe atom. (b) Brillouin zone division graph with $N=80$: the discrete green stars labeled with numbers are the crossing points of  patch center lines and  Fermi surfaces. }\label{band-patch}
\end{figure}

We adopt the band structure of the optimally doped 122-iron-pnictides.   The band structure is described by a five-orbital tight-binding model\cite{Graser2010},
\begin{equation}
H_0 = \sum_{\mathbf{k},\sigma}\sum_{a,b=1}^{5} ( \xi_{ab}(\mathbf{k}) + \epsilon_{a}\delta_{ab} )c_{a\sigma}^{\dagger}(\mathbf{k})c_{b\sigma}(\mathbf{k}),
\label{H0}\end{equation}
where $a,b$ stand for the five Fe $d$ orbitals, $\sigma$ stands for spin, $\xi_{ab}(\mathbf{k})$ is the kinetic term, $\epsilon$ is the onsite energy, $c_{a\sigma}^{\dagger}(\mathbf{k})$ creates an electron with spin $\sigma$ and momentum $\mathbf{k}$ in orbital $a$. The parameters used in Eq.\ref{H0} can be found in Graser's work\cite{Graser2010}. Throughout the rest of this paper, we take the $0.317$ hole doping.  The band structure and the  BZ division are shown  in Fig.\ref{band-patch}. There are three bands crossing Fermi level which forms five Fermi surfaces: two hole pockets centered at $(0,0)$, one hole pocket centered at $(\pi,\pi)$, and two electron pockets centered at $(\pi,0)/(0,\pi)$. More details of the model and FRG calculation can be found in our former work\cite{Yuan2014}.

% \begin{figure}%[t]
% \includegraphics[height=4cm]{dope025_U175J018.pdf}
% \caption{ (Color online) The two dominant gap functions with symmetries $A_{2g}$ (a) and $B_{1g}$ (b) at $x$=0.25, calculated with $U=1.75$ and $J=0.18$. The two pairing states are almost degenerate. }\label{UJ_0.25}
% \end{figure}

The total Hamiltonian is given by $H=H_0+H_I$, where $H_I$ describes effective electron-electron interactions. In the following sections, we will discuss the superconducting pairing symmetry under a different kind of interactions.

\section{ Onsite Repulsive Coulomb Interaction}

In this section, we concentrate on the onsite interactions which  include the intra- and inter-orbital Coulomb coupling $U$ and $U'$, Hund's coupling $J_H$, and pairing hopping term $J_{pair}$. The $H_I$ can be written as
\begin{eqnarray}
H_I & =  \sum_i[ & U\sum_a n_{ia\uparrow}n_{ia\downarrow} + U^\prime\sum_{\substack{a<b \\ \sigma,\sigma^\prime}}n_{ia\sigma}n_{ib\sigma^\prime}\nonumber \\
 &  & +\sum_{a<b} ( J_H \sum_{\sigma,\sigma^\prime}c_{ia\sigma}^{\dagger}c_{ib\sigma^\prime}^{\dagger}c_{ia\sigma^\prime}c_{ib\sigma} \nonumber \\
 & & +J_{pair}c_{ia\uparrow}^{\dagger}c_{ia\downarrow}^{\dagger}c_{ib\downarrow}c_{ib\uparrow} )],
\label{coulomb}\end{eqnarray}
where $i$ labels the site of a square lattice, $\sigma,\sigma^\prime$ label the spin, and $n_{ia\sigma}$ is number operator at site $i$ of spin $\sigma$ in orbital $a$.

\begin{figure}
  \centering
  \includegraphics[width=8.8cm]{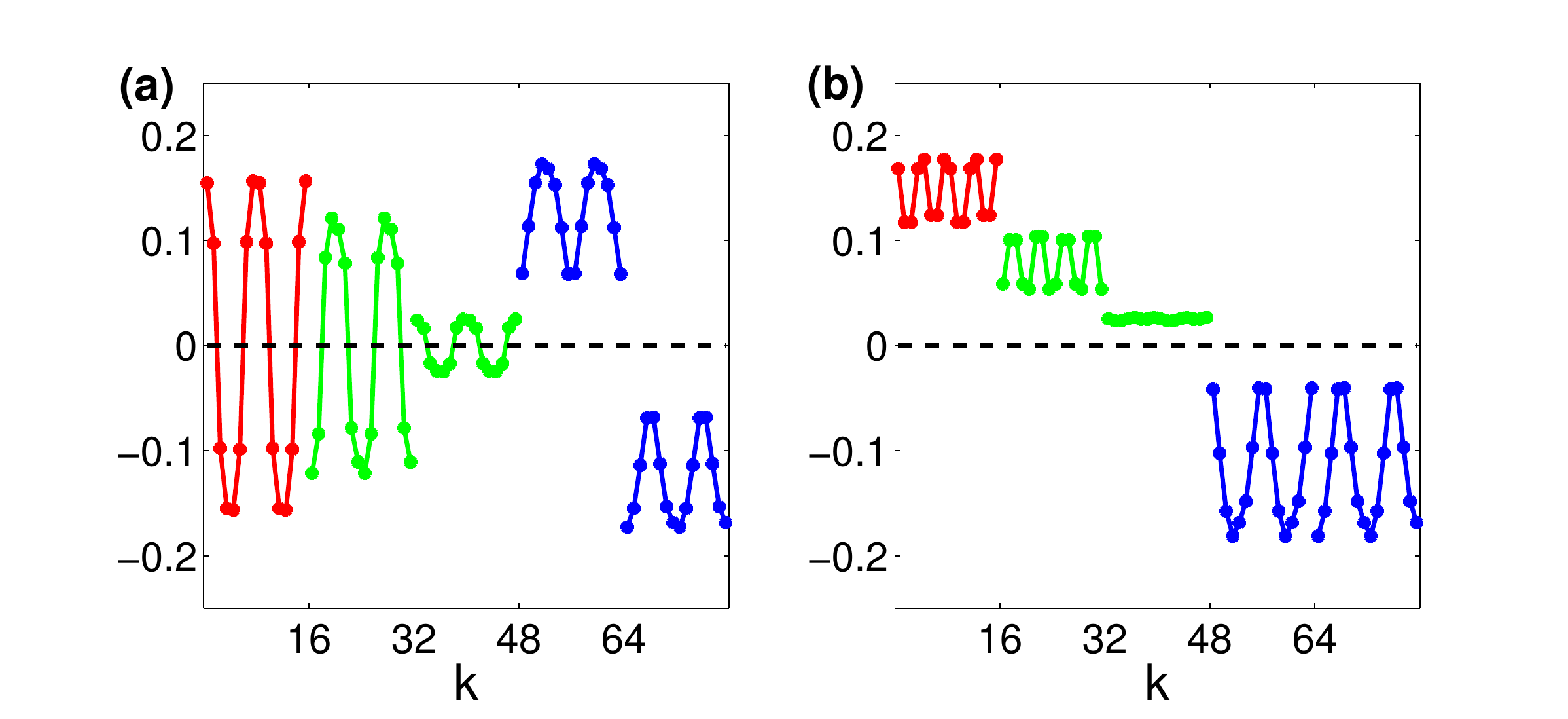}\\
  \caption{(color online)The form factor $f_1(k)$ associated to the leading superconducting instability is plotted along the five Fermi surfaces according to the numbering in Fig.\ref{band-patch}(b). The interaction parameters used in (a) are $U=3, \alpha=0.3$, and in (b) are $U=3,\alpha=0.6$.}\label{Coulomp-formfactor}
\end{figure}

Here, we maintain the basic relation $U = U'+2J_H$, $J_H = J_{pair}$ and set $J_H = \alpha U$ so that $U$ and $\alpha$ are the only two variables in this type of interaction.  Throughout this paper, we take eV as the energy unit.   In our calculation, we find that  the superconducting instability manifests into two pairing states, $d_{x^2-y^2}$-wave and $s\pm$-wave, in the parameter space.  When $U=3, \alpha = 0.3$, the system  produces $d_{x^2-y^2}$-wave and when $U=3, \alpha = 0.6$, the system  produces $s\pm$-wave, the corresponding  form factors are shown in  Fig.\ref{Coulomp-formfactor}. More detailed results are listed in Tab.\ref{coulomb-pairing}, where we can see that, when  $U\leq6$,  smaller $\alpha$ tends to $d_{x^2-y^2}$-wave  and larger $\alpha$ tends to $s\pm$-wave, and when $U>6$,  the pairing symmetry maintains to be robust $s\pm$-wave regardless of the value of $J_H$.  These results are consistent with the previous studies\cite{Thomale2011,Thomale2011-1,Wang2009-2} where $\alpha$ was taken to be large so that the $s^\pm$-wave was obtained.

\begin{table}\caption{The pairing symmetry with $U, U', J_H, J_{pair}$ interactions. The parameters satisfy $U=U'+2J_H$,$J_H = J_{pair}$ and $J_H=\alpha*U$.}
\begin{tabular}{C{1.1cm}|C{1.0cm}|C{1.0cm}|C{1.0cm}|C{1.0cm}|C{1.0cm}}
  \hline
  \hline
  ~ &  $ ~\alpha=0~$ & $\alpha=0.1$ & $\alpha=0.2$ & $\alpha=0.3$ & $\alpha=0.4$  \\
  \hline
  $~U=2$ & $d_{x^2-y^2}$ & $d_{x^2-y^2}$ & $s\pm$ & $s\pm$ & $s\pm$ \\
  \hline
  $~U=3$ & $d_{x^2-y^2}$ & $d_{x^2-y^2}$ & $s\pm$ & $s\pm$ & $s\pm$ \\
  \hline
  $~U=4$ & $d_{x^2-y^2}$ & $d_{x^2-y^2}$ & $d_{x^2-y^2}$ & $s\pm$ & $s\pm$ \\
  \hline
  $~U=5$ & $d_{x^2-y^2}$ & $d_{x^2-y^2}$ & $d_{x^2-y^2}$ & $s\pm$ & $s\pm$ \\
  \hline
  $~U=6$ & $d_{x^2-y^2}$ & $d_{x^2-y^2}$ & $d_{x^2-y^2}$ & $s\pm$ & $s\pm$ \\
  \hline
  $~U=7$ & $s\pm$ & $s\pm$ & $s\pm$ & $s\pm$ & $s\pm$ \\
  \hline
  $~U=8$ & $s\pm$ & $s\pm$ & $s\pm$ & $s\pm$ & $s\pm$ \\
  \hline
  \hline
\end{tabular}\label{coulomb-pairing}
\end{table}

\section{Effective magnetic exchange interactions}
In this section, we address pairing symmetries in a model with effective magnetic exchange interactions.
In iron-based superconductors, there are three types of magnetic exchange couplings in an effective model. The first one is the onsite Hund's couplings $J_H$(see Eq.\ref{hunds}).  The others are the NN and NNN magnetic exchange couplings, $J_1$(see Eq.\ref{J1}) and $J_2$(see Eq.\ref{J2})\cite{Bernevig2008}. Taken together, the interaction Hamiltonian can be written  as $H_I = H_{J_H}+H_{J_1}+H_{J_2}$ with
\begin{eqnarray}
H_{J_H} &=& -J_H \sum_i\sum_{a \neq b}\mathbf{S}_{ia}\cdot\mathbf{S}_{ib},
\label{hunds}\end{eqnarray}
\begin{equation}
H_{J_1} = J_1 \sum_{<i,j>}\sum_{a,b}\mathbf{S}_{ia}\cdot\mathbf{S}_{jb},
\label{J1}\end{equation}
\begin{equation}
H_{J_2} = J_2 \sum_{<<i,j>>}\sum_{a,b}\mathbf{S}_{ia}\cdot\mathbf{S}_{jb},
\label{J2}\end{equation}
where $\mathbf{S}_{ia}=\frac{1}{2}\sum_{\alpha,\beta}c_{ia\alpha}^{\dagger}\mathbf{\sigma}_{\alpha\beta}c_{ia\beta}$ is the local spin operator, $\mathbf{\sigma}$ is the Pauli matrix, $i,j$ are the lattice sites, $\alpha,\beta$ are spin indexes, and $a,b$ are orbital indexes.
%The AFM $J_1$ and $J_2$ are both positive.

Then we do Fourier transform  $c_{ia}^{\dagger}=\sum_k c_{ka}^{\dagger} e^{i \mathbf{k}\mathbf{R}_i}$ and use the relation $\mathbf{\sigma}_{\alpha\beta} \cdot \mathbf{\sigma}_{\alpha'\beta'} = 2\delta_{\alpha\beta'}\delta_{\beta\alpha'}-\delta_{\alpha\beta}\delta_{\alpha'\beta'}$. Using FRG, as described in the previous section, we study the renormalized interactions described by the four-point vertex associated with the second quantization form in momentum space,
\begin{eqnarray}
\sum_{\substack{k_1,k_2, \\ k_3,k_4}}[V_1(k_1,k_2,k_3,k_4)c_{k_1a\alpha}^{\dagger}c_{k_2b\beta}^{\dagger}c_{k_4a\beta}c_{k_3b\alpha}  \nonumber \\
+V_2(k_1,k_2,k_3,k_4)c_{k_1a\alpha}^{\dagger}c_{k_2b\beta}^{\dagger}c_{k_4b\beta}c_{k_3a\alpha}],
\end{eqnarray}
with
\begin{eqnarray}
&V_1 = -J_1(cos(k_{2x}-k_{3x})+cos(k_{2y}-k_{3y})) \nonumber \\
&- 2J_2cos(k_{2x}-k_{3x})cos(k_{2y}-k_{3y}) + \frac{1}{2}J_H,
\end{eqnarray}
and
\begin{eqnarray}
&V_2 = -\frac{1}{2}J_1(cos(k_{1x}-k_{3x})+cos(k_{1y}-k_{3y}))\nonumber \\
&-J_2cos(k_{1x}-k_{3x})cos(k_{1y}-k_{3y}) + \frac{1}{4}J_H,
\end{eqnarray}
where $k_1,k_2,k_3,k_4$ denote the momenta of the four particles. Required by the Pauli exclusion principle, $a \neq b$ for the $J_H$ term.

%\begin{widetext}
\begin{center}\begin{table*}\caption{The pairing symmetry in the presence of  the effective NN, NNN and Hund's magnetic exchange couplings.}
\begin{tabular}{C{0.9cm}|C{1.2cm}|C{1.2cm}|C{0.9cm}|C{0.9cm}|C{0.9cm}|C{0.9cm}|C{0.9cm}|C{0.9cm}|C{0.9cm}|C{0.9cm}|C{0.9cm}|C{0.9cm}}
  \hline
  \hline
   & \multicolumn{4}{|c|}{ $J_2=0$ } & \multicolumn{4}{c|}{ $J_1=0$ } & \multicolumn{4}{c}{ $J_1=J_2$ } \\
   \hline
  ~ & $J_1=0.5$ & $J_1=1.0$ & $J_1=1.5$ & $J_1=2.0$ & $J_2=0.5$ & $J_1=1.0$ & $J_2=1.5$ & $J_2=2.0$ &$J_1=0.5$ & $J_1=1.0$ & $J_1=1.5$ & $J_1=2.0$  \\
  \hline
  $J_H=0.5$ & $s\pm(nodal)$ & $s\pm(nodal)$ & $s^h_\pm$ & $s^h_\pm$ & $s\pm$ & $s\pm$ & $s\pm$ & $s\pm$  & $s\pm$ & $s\pm$ & $s\pm$ & $s\pm$ \\
  \hline
  $J_H=1.0$ & $s^h_\pm$ & $s\pm(nodal)$ & $s^h_\pm$ & $s^h_\pm$ & $s\pm$ & $s\pm$ & $s\pm$ & $s\pm$  & $s\pm$ & $s\pm$ & $s\pm$ & $s\pm$ \\
  \hline
  $J_H=1.5$ & $^h_\pm$ & $s\pm(nodal)$ & $s^h_\pm$ & $s^h_\pm$ & $s\pm$ & $s\pm$ & $s\pm$ & $s\pm$  & $s\pm$ & $s\pm$ & $s\pm$ & $s\pm$ \\
  \hline
  $J_H=2.0$ & $^h_\pm$ & $s^h_\pm$ & $s^h_\pm$ & $s^h_\pm$ & $s\pm$ & $s\pm$ & $s\pm$ & $s\pm$  & $s\pm$ & $s\pm$ & $s\pm$ & $s\pm$ \\
  \hline
  \hline
\end{tabular}\label{J1J2Jh-pairing}
\end{table*}\end{center}
%\end{widetext}

\begin{figure}
  \centering
  \includegraphics[width=8.8cm]{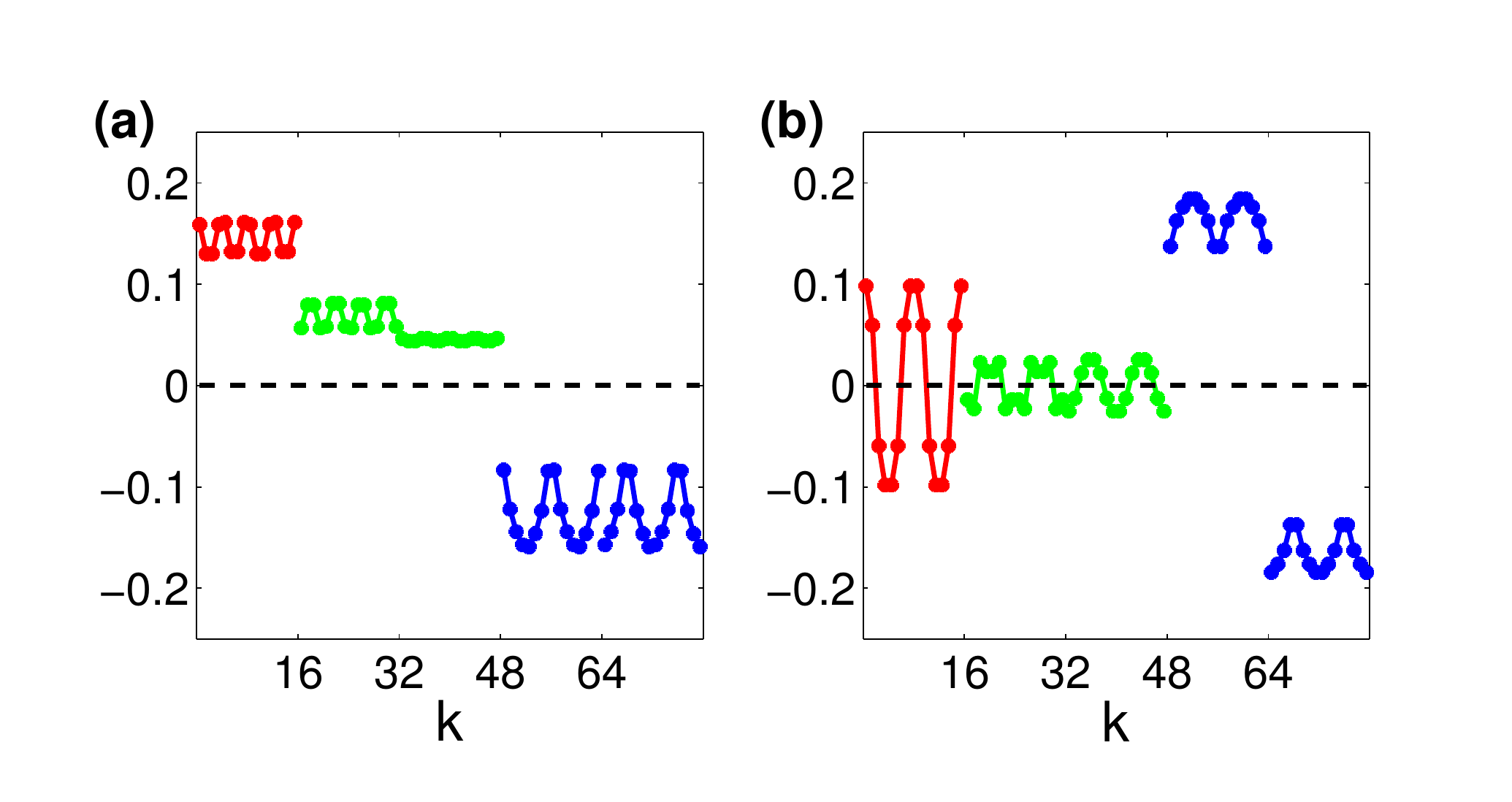}\\
  \caption{(color online)(a)The SC form factors for $J_1 = 1.0, J_2 = 1.0, and J_H = 0$: the $s\pm$-wave is robust for $J_2 \gtrsim J_1$. (b)The SC form factor for $J_1 = 3.0, J_2 = 1.0, J_H = 0$: the $d_{x^2-y^2}$ pairing state appears when $J_2 \ll J_1$.}
\label{J1J2JH-formfactor}\end{figure}

 The results are  shown in Tab.\ref{J1J2Jh-pairing} and Tab.\ref{J1J2-pairing} with various $J_1$, $J_2$, and AFM $J_H$ values. Through analyzing the results,  in this case, we  find that  the pairing symmetry is largely independent of the $J_H$ values.  Between the other two parameters, $J_2$ plays a more important role than $J_1$. It drives a $s\pm$ phase when $J_2 \gtrsim J_1$, shown in Fig.\ref{J1J2JH-formfactor}(a) and a $d_{x^2-y^2}$ state when $J_2 \ll J_1$, shown in Fig.\ref{J1J2JH-formfactor}(b). We note that in  Tab.\ref{J1J2Jh-pairing} the $s\pm(nodal)$ means that this $s\pm-wave$ has nodes in the small $\Gamma$-centered hole pocket. Furthermore, in Tab.\ref{J1J2-pairing} when $J_1=1.5, J_2=0$ and $J_1=2.0, J_2=0$, there is a  $s^h_\pm$-wave which has a sign change between the $(0,0)$-centered hole pockets and $(\pi,\pi)$-centered hole pocket or between the two $(0,0)$-centered hole pockets.

%\begin{widetext}
\begin{center}\begin{table*}\caption{The pairing symmetry with only the NN and NNN AFM exchange couplings.}
\begin{tabular}{C{1.3cm}|C{1.3cm}|C{1.3cm}|C{1.3cm}|C{1.3cm}|C{1.3cm}|C{1.3cm}|C{1.3cm}|C{1.3cm}|C{1.3cm}}
  \hline
  \hline
  $J_H=0$ & $J_2=0$ & $J_2=0.5$ & $J_2=1.0$ & $J_2=1.5$ & $J_2=2.0$ & $J_2=2.5$ & $J_2=3.0$ & $J_2=3.5$ & $J_2=4.0$ \\
   \hline
  $J_1=0$ &  & $s\pm$ & $s\pm$ & $s\pm$ & $s\pm$ & $s\pm$ & $s\pm$ & $s\pm$ &$s\pm$ \\
  \hline
  $J_1=0.5$ & $s\pm$ & $s\pm$ & $s\pm$ & $s\pm$ & $s\pm$ & $s\pm$ & $s\pm$ & $s\pm$ & $s\pm$ \\
  \hline
  $J_1=1.0$ & $s\pm$ & $s\pm$ & $s\pm$ & $s\pm$ & $s\pm$ & $s\pm$ & $s\pm$ & $s\pm$ & $s\pm$ \\
  \hline
  $J_1=1.5$ & $s^h_\pm$ & $s\pm$ & $s\pm$ & $s\pm$ & $s\pm$ & $s\pm$ & $s\pm$ & $s\pm$ & $s\pm$ \\
  \hline
  $J_1=2.0$ & $s^h_\pm$ & $s\pm$ & $s\pm$ & $s\pm$ & $s\pm$ & $s\pm$ & $s\pm$ & $s\pm$ & $s\pm$ \\
  \hline
  $J_1=2.5$ & $d_{x^2-y^2}$ & $d_{x^2-y^2}$ & $s\pm$ & $s\pm$ & $s\pm$ & $s\pm$ & $s\pm$ & $s\pm$ & $s\pm$ \\
  \hline
  $J_1=3.0$ & $d_{x^2-y^2}$ & $d_{x^2-y^2}$ & $d_{x^2-y^2}$ & $s\pm$ & $s\pm$ & $s\pm$ & $s\pm$ & $s\pm$ & $s\pm$ \\
  \hline
  $J_1=3.5$ & $d_{x^2-y^2}$ & $d_{x^2-y^2}$ & $d_{x^2-y^2}$ & $d_{x^2-y^2}$ & $s\pm$ & $s\pm$ & $s\pm$ & $s\pm$ & $s\pm$ \\
  \hline
  $J_1=4.0$ & $d_{x^2-y^2}$ & $d_{x^2-y^2}$ & $d_{x^2-y^2}$ & $d_{x^2-y^2}$ & $d_{x^2-y^2}$ & $s\pm$ & $s\pm$ & $s\pm$ & $s\pm$ \\
  \hline
  \hline
\end{tabular}\label{J1J2-pairing}
\end{table*}\end{center}
%\end{widetext}

%\begin{figure}
%  \centering
%  % Requires \usepackage{graphicx}
%  \includegraphics[width=9.5cm]{J1J2JH-formfactor(noytick).eps}\\
%  \caption{(color online)(a)The SC form factor for $J_1 = 1.0, J_2 = 1.0, J_H = 0$. It is $s\pm$ pairing. And the $s\pm$ state is robust for $J_2 \gtrsim J_1$.(b)The SC form factor for $J_1 = 3.0, J_2 = 1.0, J_H = 0$. This $d_{x^2-y^2}$ pairing state appears when $J_2 \ll J_1$.}
%\label{J1J2JH-formfactor2}\end{figure}

\section{The  orbital-dependent magnetic exchange interactions}
For the purpose of studying the interactions with greater depth, we divide the NN AFM $J_1$ and the NNN AFM $J_2$ into two parts respectively:  the intra-orbital part and  the inter-orbital part.
We first  study the interaction containing intra-orbital $J_1$  and inter-orbital $J_2$. The initial interaction Hamiltonian is
\begin{eqnarray}
J_1 \sum_{<i,j>}\sum_{a}\mathbf{S}_{ia}\cdot\mathbf{S}_{ja} + J_2 \sum_{<<i,j>>}\sum_{a \neq b}\mathbf{S}_{ia}\cdot\mathbf{S}_{jb}.
\end{eqnarray}

\begin{figure}
  \centering
  \includegraphics[width=8.2cm]{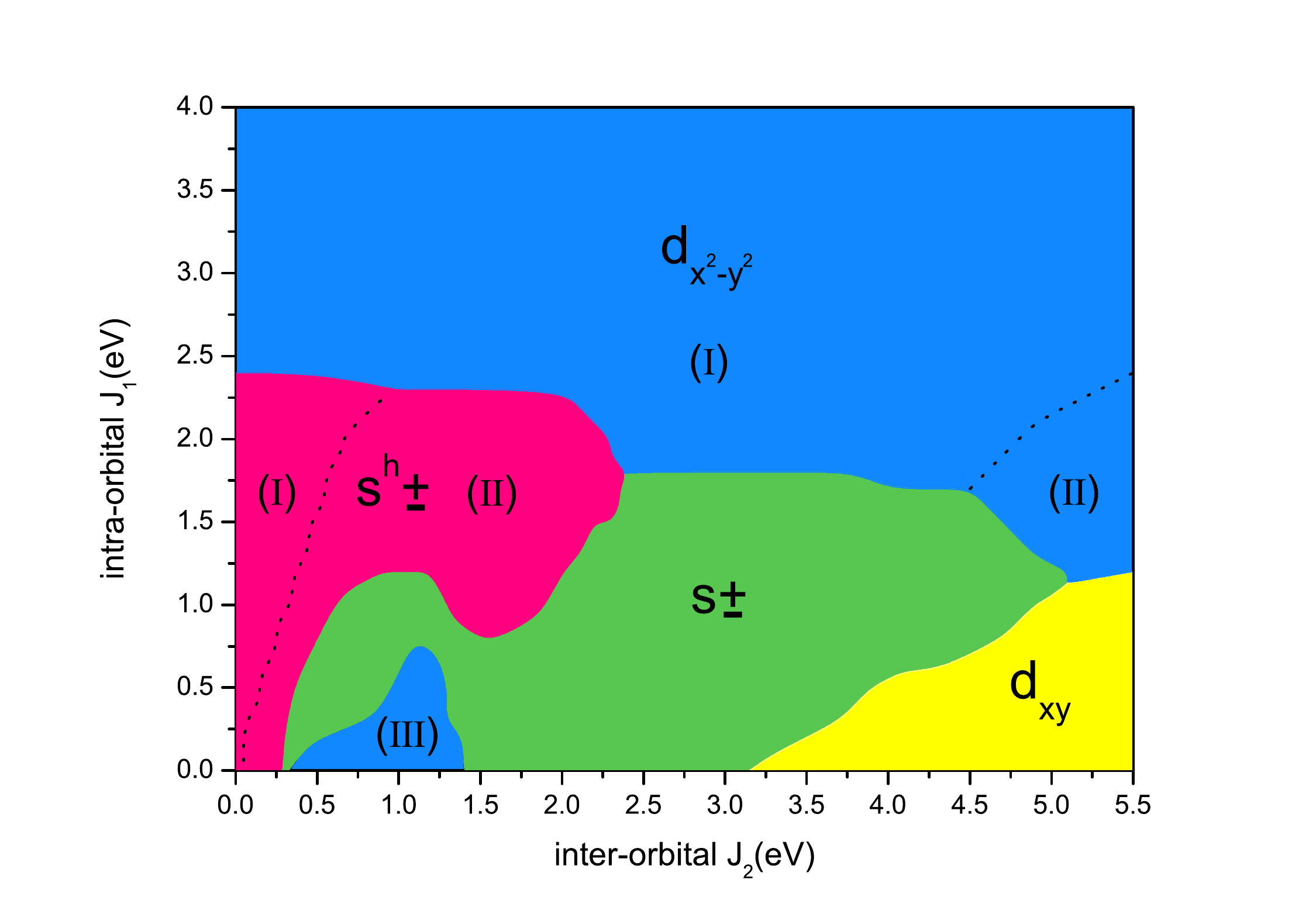}\\
  \caption{(color online)The superconducting pairing symmetry phase diagram in the intra-orbital $J_1$ and inter-orbital $J_2$ plane. When the interaction parameters vary, there are four different pairing symmetries which are shown by different colors: $s\pm$-wave (green), $d_{xy}$-wave (yellow), $d_{x^2-y^2}$-wave (blue) and $s^h_{\pm}$-wave (red).  The$d_{x^2-y^2}$-wave has three different form factors and the $s^h_{\pm}$-wave has two.  }\label{phasediagram1}
\end{figure}

 This Hamiltonian exhibits  a  more plentiful pairing phase diagram compared to the full $J_1$ and $J_2$ interaction. Varying the intra-orbital $J_1$ and the inter-orbital $J_2$ values in the FRG permitted range, we calculate the pairing symmetry numerically and get the phase diagram which is specified in Fig.\ref{phasediagram1}.

\begin{figure}
  \centering
  \includegraphics[width=8.2cm]{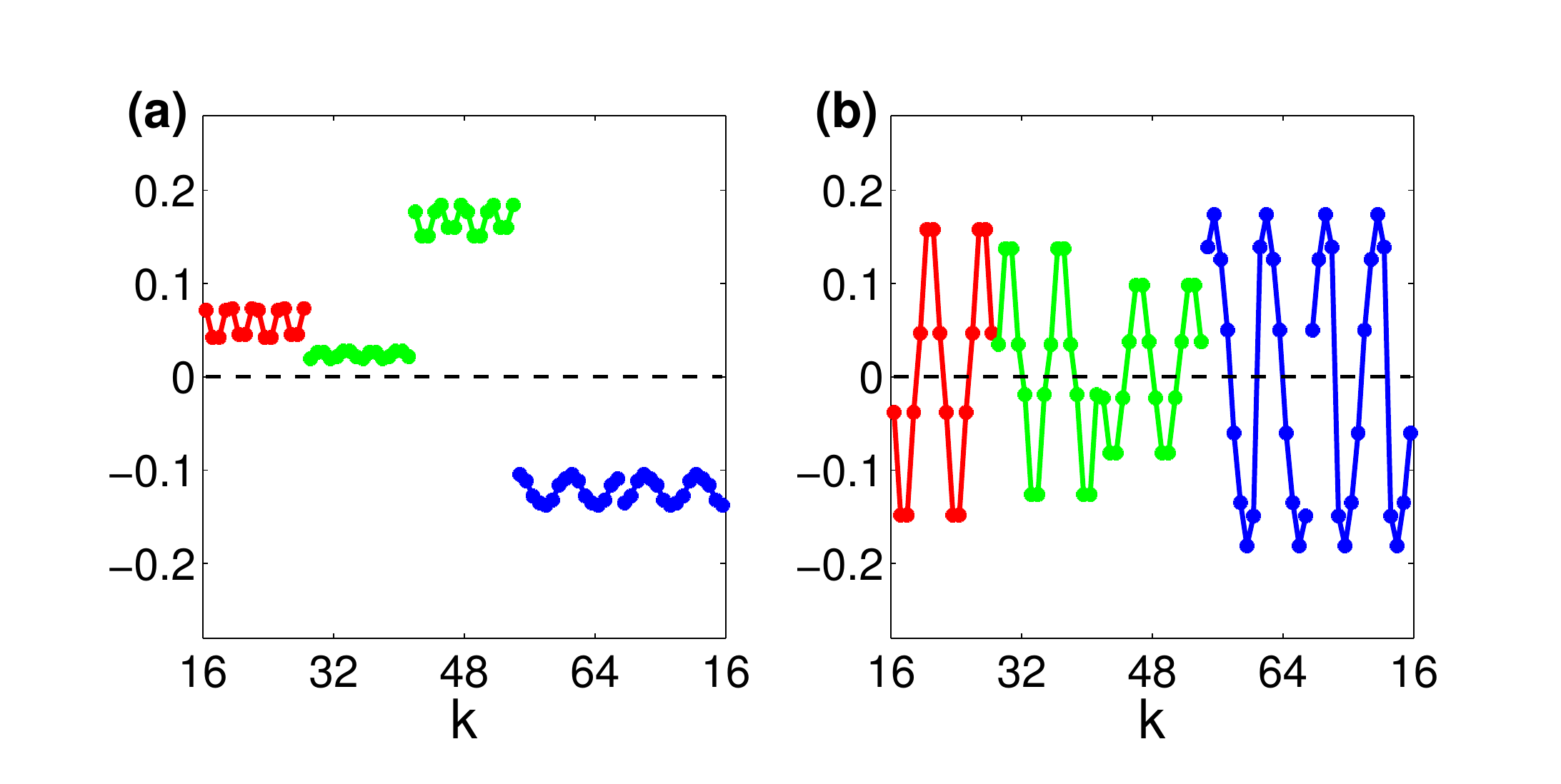}\\
  \caption{(color online) The typical SC form factors for $s\pm$-wave(a) and $d_{xy}$-wave(b) in the phase diagram Fig.\ref{phasediagram1}. The interaction parameters used in (a) are $J_1 = 1.0, J_2 = 2.5$, and in (b) $J_1 = 0.5, J_2 = 5.0$.}
\label{J1J2JH-formfactor1}
\end{figure}

\begin{figure}
  \centering
  \includegraphics[width=8.8cm]{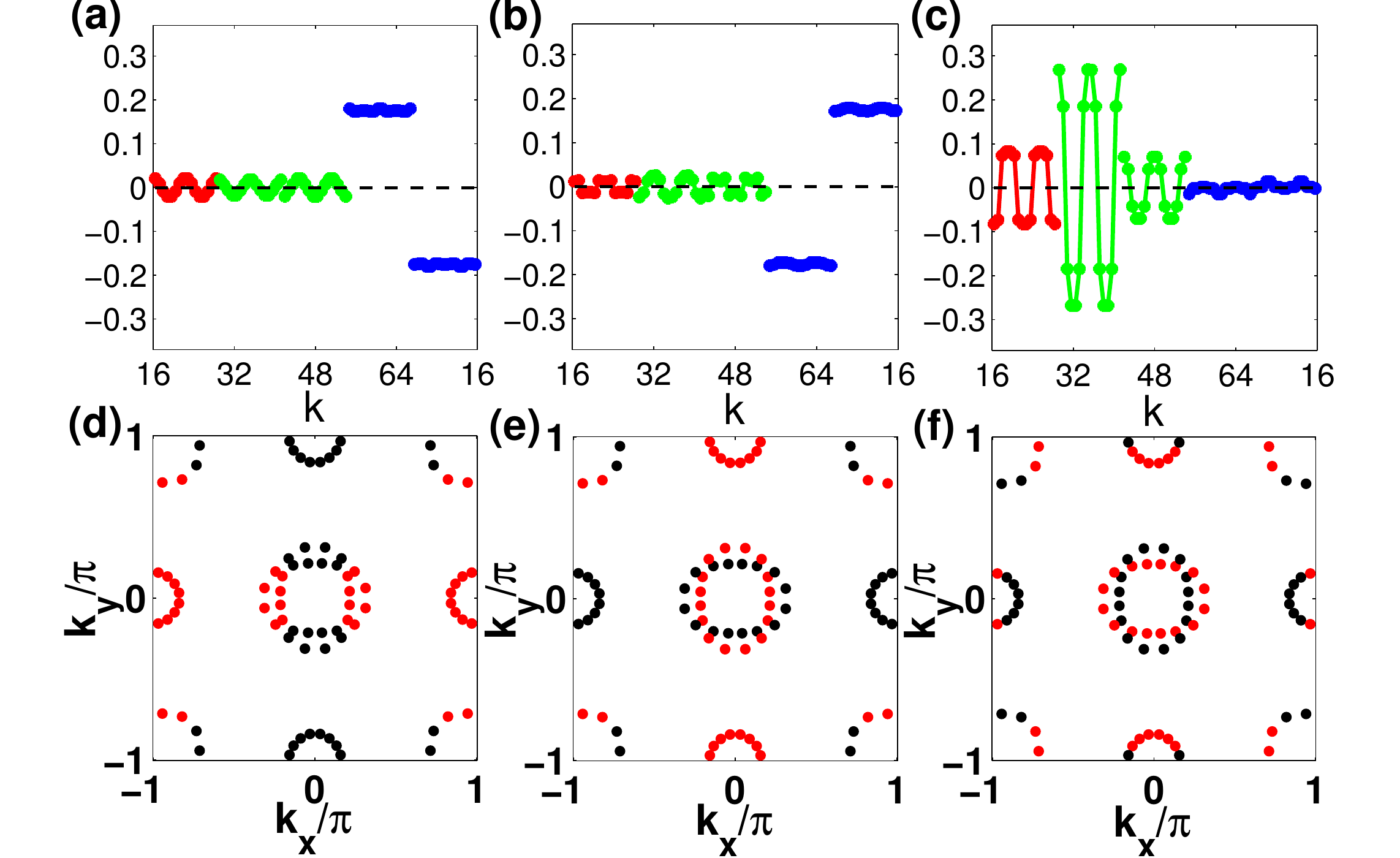}\\
  \caption{(color online) The SC form factors(a-c) and the signs of the gap functions(d-e)(the red and black points present opposite sign) for $d_{x^2-y^2}$-wave in the phase diagram(Fig.\ref{phasediagram1}). (a,d)The typical SC form factors and the corresponding signs  for the blue (I) region in Fig.\ref{phasediagram1} with $J_1 = 3.0, J_2 = 3.0$. (b,e)The typical SC form factors and the corresponding signs   for the blue (II) region in Fig.\ref{phasediagram1} with $J_1 = 1.6, J_2 = 5.0$. (c,f)The typical SC form factor and the corresponding signs  for the blue (III) region in Fig.\ref{phasediagram1} with $J_1 = 0.5, J_2 = 1.1$. }
\label{J1J2JH-formfactor2}\end{figure}

\begin{figure}
  \centering
  \includegraphics[width=6.9cm]{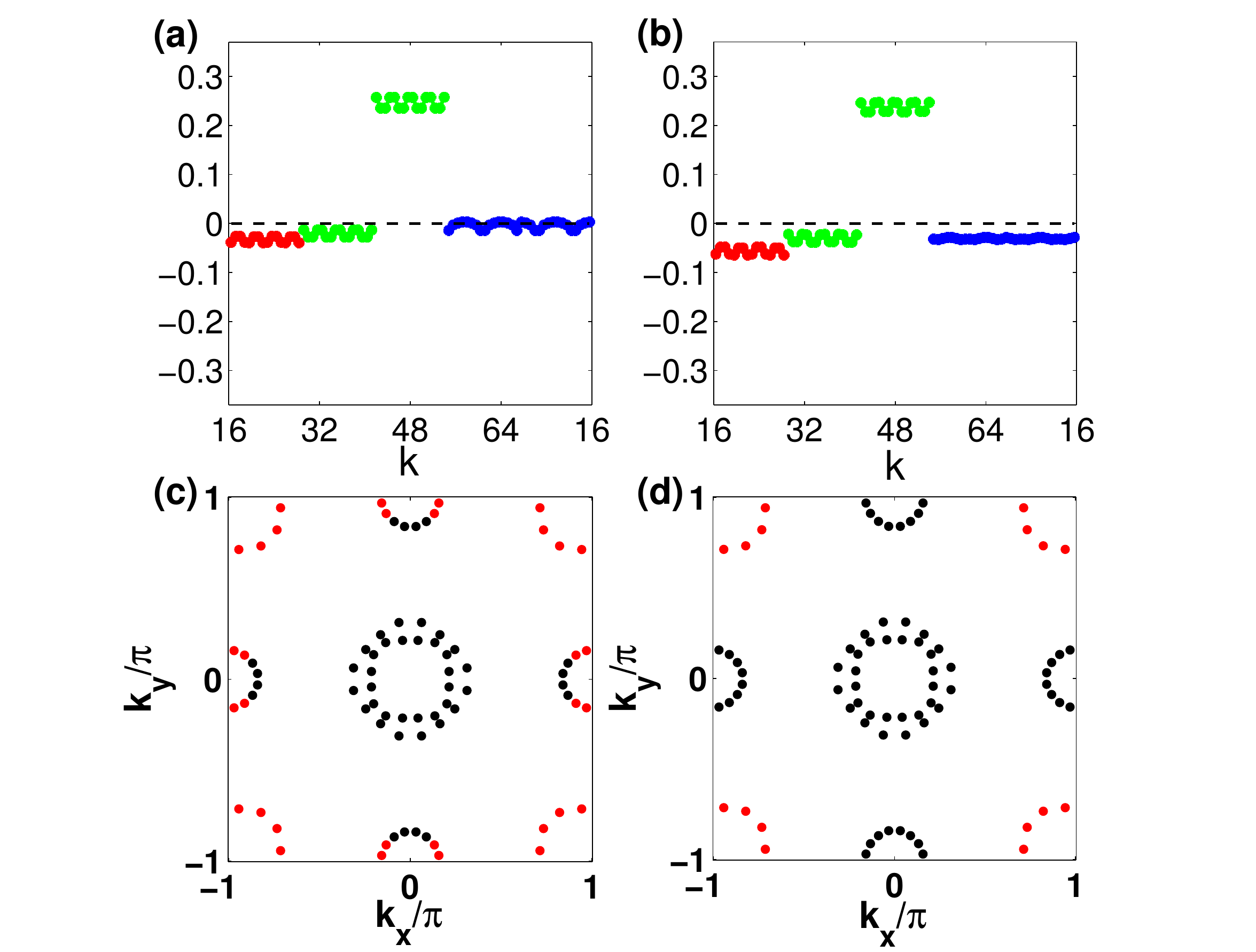}\\
  \caption{(color online) The SC form factors(a-c) and the signs(d-e) for the $s^h_{\pm}$-wave in the phase diagram(Fig.\ref{phasediagram1}). (a,c)The typical SC form factors and the corresponding signs for  the red (I) region in Fig.\ref{phasediagram1} with $J_1 = 1.5, J_2 = 0.3$. (b,d)The typical SC form factors and the corresponding signs for the red (II) region in Fig.\ref{phasediagram1} with $J_1 = 2.0, J_2 = 1.5$. }
\label{J1J2JH-formfactor3}\end{figure}

In the phase diagram, the green-filled region has the $s\pm$ pairing state(the form factor is shown in Fig.\ref{J1J2JH-formfactor1}(a)) and the yellow-filled phase in the bottom right corner has $d_{xy}$ symmetry(the form factor is shown in Fig.\ref{J1J2JH-formfactor1}(b)). The large-scaled blue-filled phase represents $d_{x^2-y^2}$ symmetry and it is subdivided into three parts which have slightly difference. The form factors of these three $d_{x^2-y^2}$ phases are shown in Fig.\ref{J1J2JH-formfactor2}. The red-filled region represents $s^h_{\pm}$ pairing symmetry where intra-orbital $J_1$  lies in $0\sim2.4$ and inter-orbital $J_2$ lies in $0\sim5.5$. In our calculation, we note that there are two  $s^h_{\pm}$  phases, the phase labeled by $s^h_{\pm}$(I)  has nodal electron pockets(the form factor is shown in Fig.\ref{J1J2JH-formfactor3}(a)) and the region $s^h_{\pm}$(II) has nodeless electron pockets(the form factor is shown in Fig.\ref{J1J2JH-formfactor3}(b)).


Secondly, we take  the interactions which  contain inter-orbital $J_1$ and the intra-orbital $J_2$. In this case, the interaction part of Hamiltonian is given by
\begin{eqnarray}
J_1 \sum_{<i,j>}\sum_{a \neq b}\mathbf{S}_{ia}\cdot\mathbf{S}_{jb} + J_2 \sum_{<<i,j>>}\sum_{a}\mathbf{S}_{ia}\cdot\mathbf{S}_{ja}.
\end{eqnarray}

\begin{figure}
  \centering
  \includegraphics[width=8.2cm]{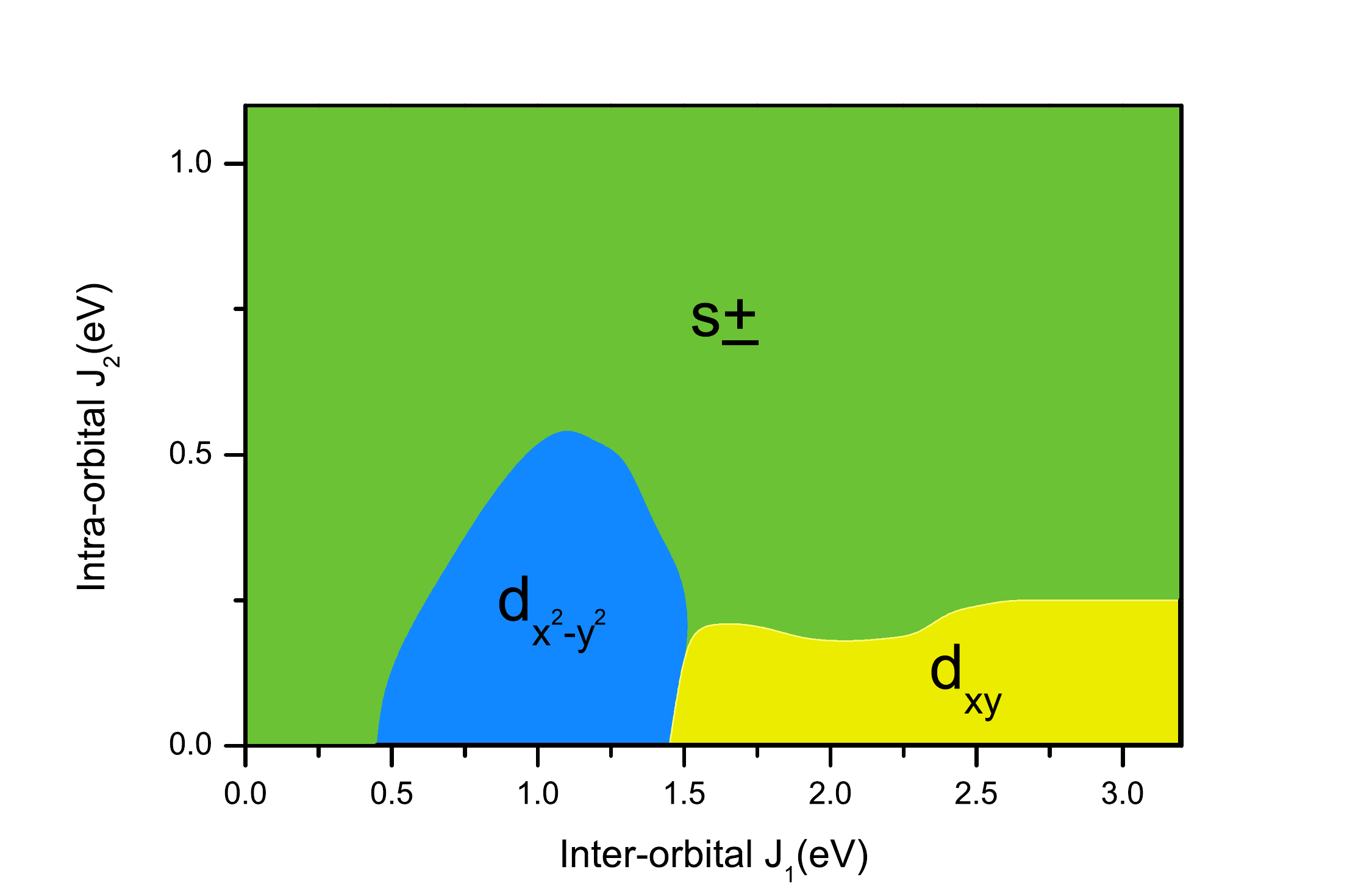}\\
  \caption{(color online)The superconducting pairing symmetry phase diagram in the inter-orbital $J_1$ and intra-orbital $J_2$ plane. There are three
   different pairing symmetry states: $s\pm$-wave, $d_{x^2-y^2}$-wave and $d_{xy}$-wave. }\label{phasediagram2}
\end{figure}

 The  result of FRG calculation is summarized in the phase diagram shown in Fig.\ref{phasediagram2}.  There are three phases in the phase diagram in the inter-orbital $J_1$ and the intra-orbital $J_2$ plane. When the intra-orbital $J_2$ is less than $0.5$ and the inter-orbital $J_1$ lies in $(0.4,1.4)$, it exhibits $d_{x^2-y^2}$ state which is in the blue region. When the intra-orbital $J_2$ is less than $0.4$ and the inter-orbital $J_1$ is greater than $1.5$, it exhibits $d_{xy}$ state, as shown in the yellow region in Fig.\ref{phasediagram2}. The remaining green region in Fig.\ref{phasediagram2} represents $s\pm$ state.

\section{Discussion and Summary}
From our calculations, it is clear that the results of FRG on  superconducting pairing symmetries  greatly depend on  the initial interactions.  Even in a fixed Fermi surface topology which was largely acknowledged to host a $s\pm$-wave  in the previous studies, a small variation of interaction parameters can lead to other pairing symmetries and  result in a complicated phase diagram. These results essentially suggest that the FRG method lacks predicting power for complex systems such as iron-based superconductors since it is difficult to extract an accurate effective model with precise estimation of interaction parameters.

Nevertheless,  robust results can be observed in our FRG results. Firstly,  the intra-orbital interaction is generally more important than the inter-orbital interaction in determining pairing symmetries. Once the intra-orbital interactions are large enough, the pairing symmetry appears to be rather robust. Secondly,  in general, the large NN intra-orbital AFM exchange coupling favors a $d_{x^2-y^2}$-wave and the large NNN intra-orbital AFM   favors a $s\pm$-wave.  Other pairing symmetries become possible only when the intra-orbital magnetic exchange couplings are small. Finally,  the fact that only  the  $d_{x^2-y^2}$-wave and  the $s\pm$-wave are obtained in the model with  onsite repulsive interactions suggests that the low energy effective magnetic exchange interactions induced by the onsite repulsive interactions can be approximated to an intra-orbital $J_1-J_2$ model.

As experimentally,   $s$-wave pairing symmetry  is rather robust throughout different families of iron-based superconductors\cite{Hu-swave},   the consistency between the experimental and  FRG results  clearly suggests that the intra-orbital AFM exchange coupling $J_2$ must be the dominant source for superconducting pairing.  Thus, it is likely that the  NN repulsive interactions must be important in iron-based superconductors as  it has been recently pointed out that it can stabilize the extended $s$-wave pairing by suppressing the pairing channel caused by the NN AFM exchange coupling $J_1$.

In summary, we have revisited the FRG results on pairing symmetries in iron-based superconductors. We show that even with a fixed Fermi surface topology, which previously was considered to support $s\pm$-wave,   different emergent pairing channels can be easily induced by varying interaction parameters. The results suggest an accurate effective model has to be built before FRG can help to predict pairing symmetries.

{\it Acknowledgement:}~~The work is supported by the Ministry of Science and Technology of China 973 program (No. 2015CB921300), National Science Foundation of China (Grant No. NSFC-1190020, 11334012), and   the Strategic Priority Research Program of  CAS (Grant No. XDB07000000).

%\begin{widetext}
%%\begin{eqnarray}
%\begin{flalign}
%H_{J_1} &= J_1 \sum_{<i,j>}\sum_{a,b}\mathbf{S}_{ia}\cdot\mathbf{S}_{jb} \nonumber \\
%&= J_1\sum_{a,b}\sum_{\alpha,\beta}\sum_{\substack{k_1,k_2 \\ k_3,k_4}}
% [(cos(k_{2x}-k_{3x})+cos(k_{2y}-k_{3y}))c_{k_1a\alpha}^{\dagger}c_{k_2b\beta}^{\dagger}c_{k_3b\alpha}c_{k_4a\beta}  \nonumber \\
% &~~~~~~~~~~~~~~~~~~~~~~+ \frac{1}{2}(cos(k_{2x}-k_{4x})+cos(k_{2y}-k_{4y}))c_{k_1a\alpha}^{\dagger}c_{k_2b\beta}^{\dagger}c_{k_3a\alpha}c_{k_4b\beta}]\delta(k_1+k_2-k_3-k_4)
%\end{flalign}
%%\end{eqnarray}
%\end{widetext}

%\paragraph{The next-nearest-neighbour inter-orbital AFM exchange coupling $J_2$}
%
%\begin{widetext}
%%\begin{eqnarray
%\begin{flalign}
%H_{J_2} &= J_2 \sum_{<<i,j>>}\sum_{a,b}\mathbf{S}_{ia}\cdot\mathbf{S}_{jb}\nonumber \\
%&= J_2\sum_{a,b}\sum_{\alpha,\beta}\sum_{\substack{k_1,k_2 \\ k_3,k_4}}
% [(2cos(k_{2x}-k_{3x})cos(k_{2y}-k_{3y})c_{k_1a\alpha}^{\dagger}c_{k_2b\beta}^{\dagger}c_{k_3b\alpha}c_{k_4a\beta})  \nonumber \\
% &~~~~~~~~~~~~~~~~~~~~~~~+ (cos(k_{2x}-k_{4x})cos(k_{2y}-k_{4y}))c_{k_1a\alpha}^{\dagger}c_{k_2b\beta}^{\dagger}c_{k_3a\alpha}c_{k_4b\beta}]\delta(k_1+k_2-k_3-k_4)
%\end{flalign}
%%\end{eqnarray}
%\end{widetext}

\end{document}